\documentclass[prd,preprintnumbers,amsmath,amssymb,nofootinbib,aps]{revtex4}

\usepackage{bm}
\usepackage{amsfonts}
\usepackage{mathrsfs}
\usepackage{graphicx}
\usepackage{amsmath}
\usepackage{float}
\usepackage{braket}

\begin{document}

\newcommand{\hhat}[1]{\hat {\hat{#1}}}
\newcommand{\pslash}[1]{#1\llap{\sl/}}
\newcommand{\kslash}[1]{\rlap{\sl/}#1}
\newcommand{\lab}[1]{}
\newcommand{\iref}[2]{}
\newcommand{\sto}[1]{\begin{center} \textit{#1} \end{center}}
\newcommand{\rf}[1]{{\color{blue}[\textit{#1}]}}
\newcommand{\eml}[1]{#1}
\newcommand{\el}[1]{\label{#1}}
\newcommand{\er}[1]{Eq.\eqref{#1}}
\newcommand{\df}[1]{\textbf{#1}}
\newcommand{\mdf}[1]{\pmb{#1}}
\newcommand{\ft}[1]{\footnote{#1}}
\newcommand{\n}[1]{$#1$}
\newcommand{\fals}[1]{$^\times$ #1}
\newcommand{\new}{{\color{red}$^{NEW}$ }}
\newcommand{\ci}[1]{}
\newcommand{\de}[1]{{\color{green}\underline{#1}}}
\newcommand{\ke}{\rangle}
\newcommand{\br}{\langle}
\newcommand{\lb}{\left(}
\newcommand{\rb}{\right)}
\newcommand{\lbk}{\left[}
\newcommand{\rbk}{\right]}
\newcommand{\blb}{\Big(}
\newcommand{\brb}{\Big)}
\newcommand{\nn}{\nonumber \\}
\newcommand{\p}{\partial}
\newcommand{\pd}[1]{\frac {\partial} {\partial #1}}
\newcommand{\cd}{\nabla}
\newcommand{\cc}{$>$}
\newcommand{\bqa}{\begin{eqnarray}}
\newcommand{\eqa}{\end{eqnarray}}
\newcommand{\bqe}{\begin{equation}}
\newcommand{\eqe}{\end{equation}}
\newcommand{\bay}[1]{\left(\begin{array}{#1}}
\newcommand{\eay}{\end{array}\right)}
\newcommand{\eg}{\textit{e.g.} }
\newcommand{\ie}{\textit{i.e.}, }
\newcommand{\iv}[1]{{#1}^{-1}}
\newcommand{\st}[1]{|#1\ke}
\newcommand{\at}[1]{{\Big|}_{#1}}
\newcommand{\zt}[1]{\texttt{#1}}
\newcommand{\non}{\nonumber}
\newcommand{\m}{\mu}
\newcommand{\sgn}{\operatorname{sgn}}
\newcommand\blfootnote[1]{%
  \begingroup
  \renewcommand\thefootnote{}\footnote{#1}%
  \addtocounter{footnote}{-1}%
  \endgroup
}
\renewcommand*{\thefootnote}{\dagger{footnote}}
\def\xa{{\alpha}}
\def\xA{{\Alpha}}
\def\xb{{\beta}}
\def\xB{{\Beta}}
\def\xd{{\delta}}
\def\xD{{\Delta}}
\def\xe{{\epsilon}}
\def\xE{{\Epsilon}}
\def\xve{{\varepsilon}}
\def\xg{{\gamma}}
\def\xG{{\Gamma}}
\def\xk{{\kappa}}
\def\xK{{\Kappa}}
\def\xl{{\lambda}}
\def\xL{{\Lambda}}
\def\xo{{\omega}}
\def\xO{{\Omega}}
\def\xvp{{\varphi}}
\def\xs{{\sigma}}
\def\xS{{\Sigma}}
\def\xt{{\theta}}
\def\xvt{{\vartheta}}
\def\xT{{\Theta}}
\def \Tr {{\rm Tr}}
\def\CA{{\cal A}}
\def\CC{{\cal C}}
\def\CD{{\cal D}}
\def\CE{{\cal E}}
\def\CF{{\cal F}}
\def\CH{{\cal H}}
\def\CJ{{\cal J}}
\def\CK{{\cal K}}
\def\CL{{\cal L}}
\def\CM{{\cal M}}
\def\CN{{\cal N}}
\def\CO{{\cal O}}
\def\CP{{\cal P}}
\def\CQ{{\cal Q}}
\def\CR{{\cal R}}
\def\CS{{\cal S}}
\def\CT{{\cal T}}
\def\CV{{\cal V}}
\def\CW{{\cal W}}
\def\CY{{\cal Y}}
\def\BC{\mathbb{C}}
\def\BR{\mathbb{R}}
\def\BZ{\mathbb{Z}}
\def\sA{\mathscr{A}}
\def\sB{\mathscr{B}}
\def\sF{\mathscr{F}}
\def\sG{\mathscr{G}}
\def\sH{\mathscr{H}}
\def\sJ{\mathscr{J}}
\def\sL{\mathscr{L}}
\def\sM{\mathscr{M}}
\def\sN{\mathscr{N}}
\def\sO{\mathscr{O}}
\def\sP{\mathscr{P}}
\def\sR{\mathscr{R}}
\def\sQ{\mathscr{Q}}
\def\sS{\mathscr{S}}
\def\sX{\mathscr{X}}

\def\slz{SL(2,Z)}
\def\slr{$SL(2,R)\times SL(2,R)$ }
\def\ads{${AdS}_5\times {S}^5$ }
\def\adst{${AdS}_3$ }
\def\sun{SU(N)}
\def\ad#1#2{{\frac \delta {\delta\sigma^{#1}} (#2)}}
\def\bqf{\bar Q_{\bar f}}
\def\nf{N_f}
\def\sunf{SU(N_f)}

\def\dcirc{{^\circ_\circ}}

\author{Morgan H. Lynch}
\email{mhlynch@uwm.edu}
\affiliation{Leonard E. Parker Center for Gravitation, Cosmology and Astrophysics, Department of Physics, University of Wisconsin-Milwaukee,
P.O.Box 413, Milwaukee, Wisconsin USA 53201}

\title{A Theory of Accelerated Quantum Dynamics\footnote{Essay written for the Gravity Research Foundation 2015 Awards for Essays on Gravitation.}}
\date{\today}

\begin{abstract}
The role of acceleration in particle physics can provide an alternative method for probing the properties of quantum gravity. To analyze acceleration-induced processes one utilizes the formalism of quantum field theory in curved spacetime. This quantum theory of fields in classical general relativistic backgrounds has already provided the first insights into the quantum effects of general relativity. By utilizing this formalism to compute acceleration-induced particle physics processes, we can better establish how the dynamics of elementary particles change in non-Minkowskian spacetimes. To analyze these processes, we present a theory of Accelerated Quantum Dynamics (AQD) along with certain observables predicted by the theory.
\end{abstract}

\maketitle

\section{Introduction}
The first insight into quantum gravity comes from the three canonical particle production mechanisms of quantum field theory in curved spacetime: the Parker [1], Hawking [2], and Unruh [3] effects. Inherent to each is the presence of a horizon which emits a thermal bath of particles. The Unruh effect, which is generated by the apparent horizon of an accelerated reference frame, can be studied by analyzing transitions of an accelerated two level system, i.e. a detector. The generalization of an accelerated detector undergoing a transition in energy to an accelerated particle decaying was first studied by Mueller [4]. There, the decay rate of accelerated muons, pions, and protons was computed using scalar fields. Matsas and Vanzella [5-7] further generalized the decay of protons, and similar processes, to fermions using a more appropriate V-A weak interaction while also analytically computing the power emitted by the daughter products. The spectra of emitted particles was also numerically computed. The previously analyzed decay processes were, depending on the accelerated particle, restricted to final states of the only two or three particles. In [8] we used massless scalar fields to generalize the decay process to final states of arbitrary multiplicity. This $n$-particle formalism fully extended the previous results to encompass all possible decay pathways for any accelerated particle. It was then used to compute the acceleration-induced decay rate and lifetime of electrons and muons along with the evolution of the muon decay branching fractions under acceleration. We later incorporated the emitted power, analytically computed the spectra and displacement law of the emitted particles, and included time-dependent accelerations into the formalism. This theory of particle decay induced by acceleration we call AQD [9]. It is an analytic computational framework capable of computing a wide class of closed form observables associated with the acceleration-induced decay of any accelerated particle into any final state. This enables the inclusion of high acceleration into the analysis of particle physics in regimes such as inflationary epoch of the early universe or perhaps even evaporating black holes. The following sections outline the basic formalism along with certain observables of interest.

\section{The Transition Amplitude}
To model any decay process, we consider an initial Rindler particle undergoing time-dependent hyperbolic motion along an accelerated trajectory and transitioning into a final state of arbitrary multiplicity. We let $n_{R}$ and $n$ be the number of final state Rindler and Minkowski particles respectively. The Rindler fields $\Psi_{j}$ may have arbitrary mass and are used to describe the elementary particles that are under acceleration while the Minkowski fields $\phi_{k}$ are assumed to be massless. The processes under consideration can be written as

\bqa
\Psi_{i} \rightarrow_{a} \Psi_{1} + \Psi_{2} + \cdots + \Psi_{n_{R}} + \phi_{1} + \phi_{2} + \cdots + \phi_{n}.
\eqa

That is, an initially accelerated particle decays into $n_{R}$ accelerated particles and $n$ inertial particles. In order to analyze the above arbitrary process we utilize the following action in the interaction picture. Hence,

\bqe
\hat{S}_{I} = G_{n}\int d^{4}x\sqrt{-g}\hat{\Psi}_{i}\prod_{r = 1}^{n_{R}}\hat{\Psi}_{r} \prod_{m = 1}^{n}\hat{\phi}_{m}.
\eqe

Here the coupling $G_{n}$, which is different for each multiplicity, is fixed by matching the observable under consideration to the associated process in the inertial limit. The Fock space of states for our Rindler particles $\ket{\Psi_{j}}$ are labeled by the index $j$ which characterizes their energies. We restrict the domain of the accelerated particles to the right Rindler wedge. Our Minkowski states $\ket{\mathbf{k}_{\ell}}$ are characterized by their momenta as usual and are built up from the Poincare invariant vacuum. The shorthand notation $\ket{\prod_{i}^{n}\mathbf{k}_{i}} = \ket{\mathbf{k}_{1},\mathbf{k}_{2},\cdots \mathbf{k}_{n}}$ is used to denote the Rindler and Minkowski states in a more compact fashion. The transition amplitude which governs all acceleration-induced processes is then given by

\bqe
\mathcal{A} = \bra{\prod_{\ell = 1}^{n}\mathbf{k}_{\ell}}\otimes \bra{\prod_{j = 1}^{n_{R}}\Psi_{j}} \hat{S}_{I}\ket{\Psi_{i}} \otimes \ket{0}.
\eqe

Here we have an initial Minkowski vacuum state and an initial Rindler state with one accelerated particle and, after the interaction, a final Minkowski state and final Rindler state of arbitrary multiplicity. From this amplitude we can compute various observables associated with this process utilizing a generalization of Fermi's golden rule to encompass time-dependent accelerations. Among the most important observables are the transition rate, multiplicity, emitted power, final state spectra, and displacement law. These will be presented in the following sections. Complete derivations of all observables are presented in [9]. For an acceleration $a$ and a total change in Rindler space energy $\Delta E$, the transition amplitude is diagrammatically represented by the spacetime diagram in Fig. 1.

\begin{figure}[H]
\centering  
\includegraphics[,scale=1]{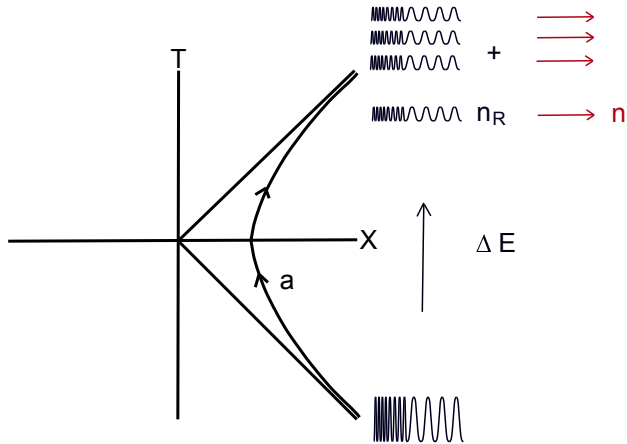}
\caption{A diagrammatic representation of the acceleration-induced transition.}
\end{figure}

\section{Transition Rate}

The acceleration-induced transition rate is characterized by an integer indexed polynomial of multiplicity along with a thermal distribution of bosonic statistics. The $n=1$ case reproduces the standard Unruh effect computation. The transition rates up and down in energy are related to each other by the detailed balance of processes at thermal equilibrium. The lifetime is determined by the reciprocal of the transition rate. This also implies the lifetime of a particle, be it stable or unstable, is acceleration-dependent. Figure 2 contains plots of the transition rate for the multiplicities $n=2$, $3$, and $4$. 

\bqa
\Gamma_{n}(\Delta E, a) = G_{n}^{2}\lb\frac{\Delta E}{\pi} \rb^{2n-1} \frac{1}{(4n -2)!!} \prod_{k = 0}^{n-1}\lbk 1  + k^{2} \lb \frac{ a}{\Delta E} \rb^{2} \rbk  \frac{1}{e^{2\pi\Delta E/|a|} - 1}.
\eqa

\begin{figure}[H]
\centering  
\includegraphics[,scale=1.1]{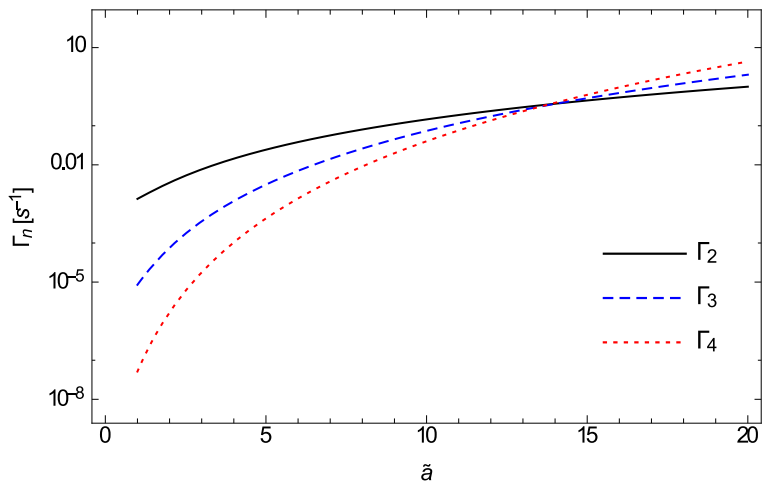}
\caption{The transition rate with $\tilde{a} = |a|/\Delta E$, $\Delta E<0$, and $G_{n} = 1$.}
\end{figure}

\section{Multiplicity}

It is clear from the transition rate plots that the dominant decay pathway is acceleration-dependent. By knowing, or perhaps even controlling, the acceleration scale of the system one would then be able to predict, or even tune, the final state multiplicity. For a given multiplicity, the inertial decay rate $\lambda_{n}$ can be found by taking the limit $a \rightarrow 0$ of Eqn. (4). Thus,

\bqa
\lambda_{n} &=& G^{2}_{n} \lb \frac{\Delta E}{\pi} \rb^{2n - 1} \frac{1}{(4n - 2)!!}.
\eqa

Moreover, this also allows us to fix the arbitrary coupling $G_{n}$ for the appropriate process in terms of the inertial rate. For two different final state multiplicities $n$ and $m$, we note $\lambda_{m}/\lambda_{n} = Br_{m}/Br_{n}$, where $Br_{i}$ is the branching fraction of the $i$th decay pathway [10]. The crossover scale is then determined by $\Gamma_{n} = \Gamma_{m}$. If we assume $n>m$, the acceleration scale where the transition rates meet can then be determined from

\bqa
\prod_{k = m}^{n - 1} \lbk 1+k^{2}\lb \frac{a}{\Delta E} \rb ^{2}\rbk &=& \frac{Br_{m}}{Br_{n}}.
\eqa

\section{Power Radiated}

The power radiated away by each of the $n$ particles during the decay process is characterized by a half integer indexed polynomial of multiplicity along with a thermal distribution of fermionic statistics. The $n = 1$ case reproduces the appropriate $a^2$ dependence for bremsstrahlung provided one makes the identification $\Delta E = 0$ as computed in [11]. Figure 3 contains plots of the emitted power for the multiplicities $n=2$, $3$, and $4$. 

\bqe
\mathcal{S}_{n}(\Delta E, a) = G^{2}_{n}4\pi \lb \frac{\Delta E}{\pi} \rb^{2n} \frac{1}{(4n)!!}  \prod_{k = 0}^{n-1}\lbk 1 + \lb \frac{2k+1}{2} \rb^{2} \lb \frac{a}{\Delta E} \rb^{2}\rbk  \frac{1}{e^{2\pi\Delta E/|a|} + 1}.
\eqe

\begin{figure}[H]
\centering  
\includegraphics[,scale=1.1]{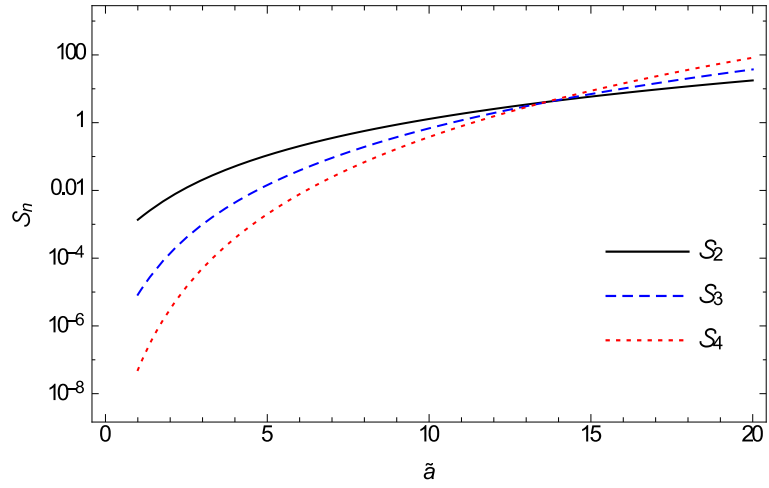}
\caption{The emitted power with $\tilde{a} = |a|/\Delta E$, $\Delta E<0$, and $G_{n} = 1$.}
\end{figure}

\section{Energy Spectra}

The energy spectra of each of the emitted $n$ particles during the decay process is characterized by an integer indexed polynomial of multiplicity along with a thermal distribution of bosonic statistics at finite chemical potential. The $n = 1$ case yields the same result as Fermi's golden rule,

\bqa
\mathcal{N}_{1}(\tilde{\omega}) = G_{1}^{2} \frac{\tilde{\omega}}{2 \pi} \delta(\Delta E+\tilde{\omega}).
\eqa

For multiplicities $n\geq 2$ the spectra are Planck-like and the change in Rindler space energy $\Delta E$ plays the role of the chemical potential. Figure 4 contains plots of the spectra, normalized to unity, for the multiplicities $n=2$, $3$, and $4$. 

\bqa
\mathcal{N}_{n}(\tilde{\omega}) &=&  \frac{G^{2}_{n}\tilde{\omega}}{(2 \pi)^{2}} \lb \frac{\Delta E+\tilde{\omega}}{\pi}\rb^{2n-3} \frac{1}{(4n-6)!!}  \prod_{k = 0}^{n-2}\lbk 1  + k^{2} \lb \frac{ a}{\Delta E+\tilde{\omega}} \rb^{2} \rbk  \frac{1}{e^{2\pi(\Delta E+\tilde{\omega})/|a|} - 1}.
\eqa

\begin{figure}[H]
\centering  
\includegraphics[,scale=1.1]{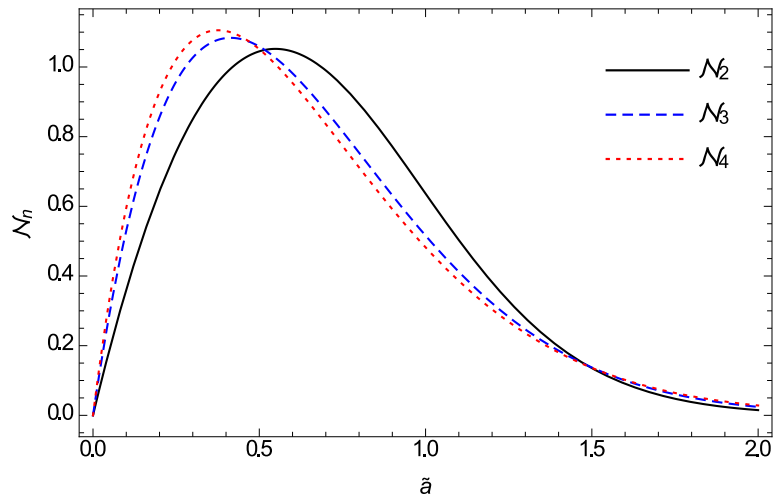}
\caption{The spectra normalized to unity with $a=1$, $\Delta E=-1$, and $G_{n} = 1$.}
\end{figure}

\section{Displacement Law}

The most probable energy of each emitted particle can be computed via the standard prescription of taking the derivative of the spectra and setting it to zero. The resulting generalized displacement law, which numerically defines the displacement coefficient $x$, is given by

\bqa
\frac{xe^{x}}{e^{x} - 1} - \lbk \frac{1}{1-\frac{2 \pi \Delta E}{|a|x}} +(2n - 3)-2\lb \frac{2 \pi}{x}\rb^{2} \sum_{k = 0}^{n-2}\frac{k^{2}}{1  + k^{2} \lb \frac{2 \pi}{x} \rb^{2}}  \rbk &=& 0.
\eqa

Once the displacement coefficient is known, the most probable energy of the emitted particle is given by

\bqe
\tilde{\omega} = x \frac{|a|}{2 \pi} - \Delta E.
\eqe

We see the peak energy follows the standard Wien's displacement law for the accelerated temperature $t_{a} = \frac{a}{2\pi}$, i.e. $\tilde{\omega} = xt_{a}$, in the limit $\Delta E = 0$. We may also define the inherently quantum mechanical energy of acceleration $E_{a}$. Upon reinsertion of the relevant physical constants we have

\bqe
E_{a} = \frac{xa\hbar}{2 \pi c}.  
\eqe

\section{Conclusions}
The formalism of AQD offers a compact set of equations which can be used to analyze acceleration-induced particle physics processes. This enables the use of particles to probe the properties of strongly accelerated or, via the equivalence principle, gravitating systems. Classes of experimentally controllable settings which exhibit high accelerations, or analogous curved spacetimes, have been discussed in [12,13] and others. The dynamics of particles during the inflationary epoch of the early universe will also undoubtedly be affected by the exponentially large accelerations [14]. The utility of AQD is that it can compute the decay rate, multiplicity, emitted power, and spectra of these highly accelerated particle decays for virtually any particle under consideration.

\section*{Acknowledgments}
The author is indebted to Luiz da Silva and Daniel Agterberg for numerous discussions. This research was funded by the Leonard E. Parker Center for Gravitation, Cosmology, and Astrophysics and the University of Wisconsin-Milwaukee Department of Physics.

\goodbreak

\end{document}